\documentclass[superscriptaddress,twocolumn,showpacs,preprintnumbers,amsmath,amssymb,prl]{revtex4-2}
\usepackage{graphicx}
\usepackage{dcolumn}
\usepackage{bm}
\usepackage{float}
\usepackage{color}
\usepackage{amsmath}
\usepackage{amssymb}
\usepackage{bm}
\usepackage{mathrsfs}
\usepackage{esint}
\usepackage{gensymb}
\usepackage{bbold}
\usepackage[unicode=true,pdfusetitle,bookmarks=true,
bookmarksnumbered=false,bookmarksopen=false,breaklinks=false,
pdfborder={0 0 0},backref=false,colorlinks=true,citecolor=red]
{hyperref}
\usepackage{cleveref}
\usepackage{comment}
\newcommand{\beginsupplement}{%
        \setcounter{table}{0}
        \renewcommand{\thetable}{S\arabic{table}}%
        \setcounter{figure}{0}
        \renewcommand{\thefigure}{S\arabic{figure}}%
     }

\allowdisplaybreaks[3]

\makeatletter
\newcommand{\manuallabel}[2]{\def\@currentlabel{#2}\label{#1}}
\makeatother

\begin{document}

\title{First order alignment transition in an interfaced active nematic}

\author{Olga Bantysh}
\affiliation{Department of Materials Science and Physical Chemistry, Universitat de Barcelona, 08028 Barcelona, Spain}
\affiliation{Institute of Nanoscience and Nanotechnology, IN2UB, Universitat de Barcelona, 08028 Barcelona, Spain}
\author{Berta Mart\'{\i}nez-Prat}
\affiliation{Department of Materials Science and Physical Chemistry, Universitat de Barcelona, 08028 Barcelona, Spain}
\affiliation{Institute of Nanoscience and Nanotechnology, IN2UB, Universitat de Barcelona, 08028 Barcelona, Spain}
\author{Jyothishraj Nambisan}
\affiliation{Department of Condensed Matter Physics, Universitat de Barcelona, 08028 Barcelona, Spain}
\author{Alberto Fern\'andez-Nieves}
\affiliation{Department of Condensed Matter Physics, Universitat de Barcelona, 08028 Barcelona, Spain}
\affiliation{University of Barcelona Institute of Complex Systems, UBICS, Universitat de Barcelona, 08028 Barcelona, Spain}
\affiliation{ICREA-Institucio Catalana de Recerca i Estudis Avan\c{c}ats, Barcelona, Spain}
\author{Francesc Sagu\'es}
\author{Jordi Ign\'es-Mullol}
\affiliation{Department of Materials Science and Physical Chemistry, Universitat de Barcelona, 08028 Barcelona, Spain}
\affiliation{Institute of Nanoscience and Nanotechnology, IN2UB, Universitat de Barcelona, 08028 Barcelona, Spain}
\email{f.sagues@ub.edu}

\date{\today}

\begin{abstract}

We investigate experimentally the dynamic phase transition of a two-dimensional active nematic layer interfaced with a passive liquid crystal. Under a temperature ramp that leads to the transition of the passive liquid into a highly anisotropic lamellar smectic-A phase, and in the presence of a magnetic field, the coupled active nematic reorganizes its flow and orientational patterns from the turbulent into a quasi-laminar regime aligned perpendicularly to the field. Remarkably, while the phase transition of the passive fluid is known to be continuous, or second order, our observations reveal intermittent dynamics of the order parameter and the coexistence of aligned and turbulent regions in the active nematic, a signature of discontinuous, or first order, phase transitions.

\end{abstract}
\maketitle

The concept of phase transition is ubiquitous in classical condensed matter physics.
It is also commonplace in active matter \cite{Ramaswamy10,marchetti13,sagues22} where, for instance, the concept of motility induced phase separation \cite{cates15} has become a paradigm to explain the aggregation of self-motile colloids in the absence of attractive forces.
By referring to dry active systems, the spontaneous assembling of flocks that coexist with a rarefied phase can be similarly understood as a phase transition that occurs in dense assembly of moving units that interact through noisy rules \cite{vanderlinden19,geyer19,chate20}.

In the context of active fluids, where the concept of phase is often used \cite{cates10,schwarz12,blow14,soni19,adkins22}, phase transitions have not been similarly pinpointed. This is remarkable since the constituents of some of their more celebrated experimental realizations display distinctive orientational symmetries, thus allowing the observed textures to be characterized in terms of standard order parameters. Particularly representative is the case of an active
nematic (AN) film, i.e., the two-dimensional aligned phase of a system of bundled microtubules crosslinked with kinesin motors, and prepared at the water/oil interface \cite{sanchez12,Doost18}.
Apart from the obvious adenosine triphosphate (ATP) concentration that fuels activity, there does not seem to be any other system parameter that can be easily identified as controlling the intrinsic chaotic flows that appear in typical unconstrained preparations \cite{giomi15,urzay17,martinez19}.

New control opportunities appear if one considers an indirect actuation mediated by the oily phase in contact with the active film.
For instance, mechanical contact of the AN with a Smectic A (SmA) liquid crystal layer \cite{oswald2006} promotes the orientation of the active flows along the low viscosity directions of the highly anisotropic interface \cite{Guillamat16,Guillamat17}.
In this Letter, we describe an order/disorder dynamic transition of the kinesin/tubulin AN between a turbulent regime and a pattern of directed, quasi-laminar flows following a temperature ramp where the oil layer alternates between a disordered nematic and a magnetically-aligned SmA phase.
We show that, while the phase transition of the passive liquid crystal is known to be second order, the concurrent alignment transition of the AN bears signatures of first order. \\

\begin{figure}[t]
	\centering
	\includegraphics[width=\linewidth]{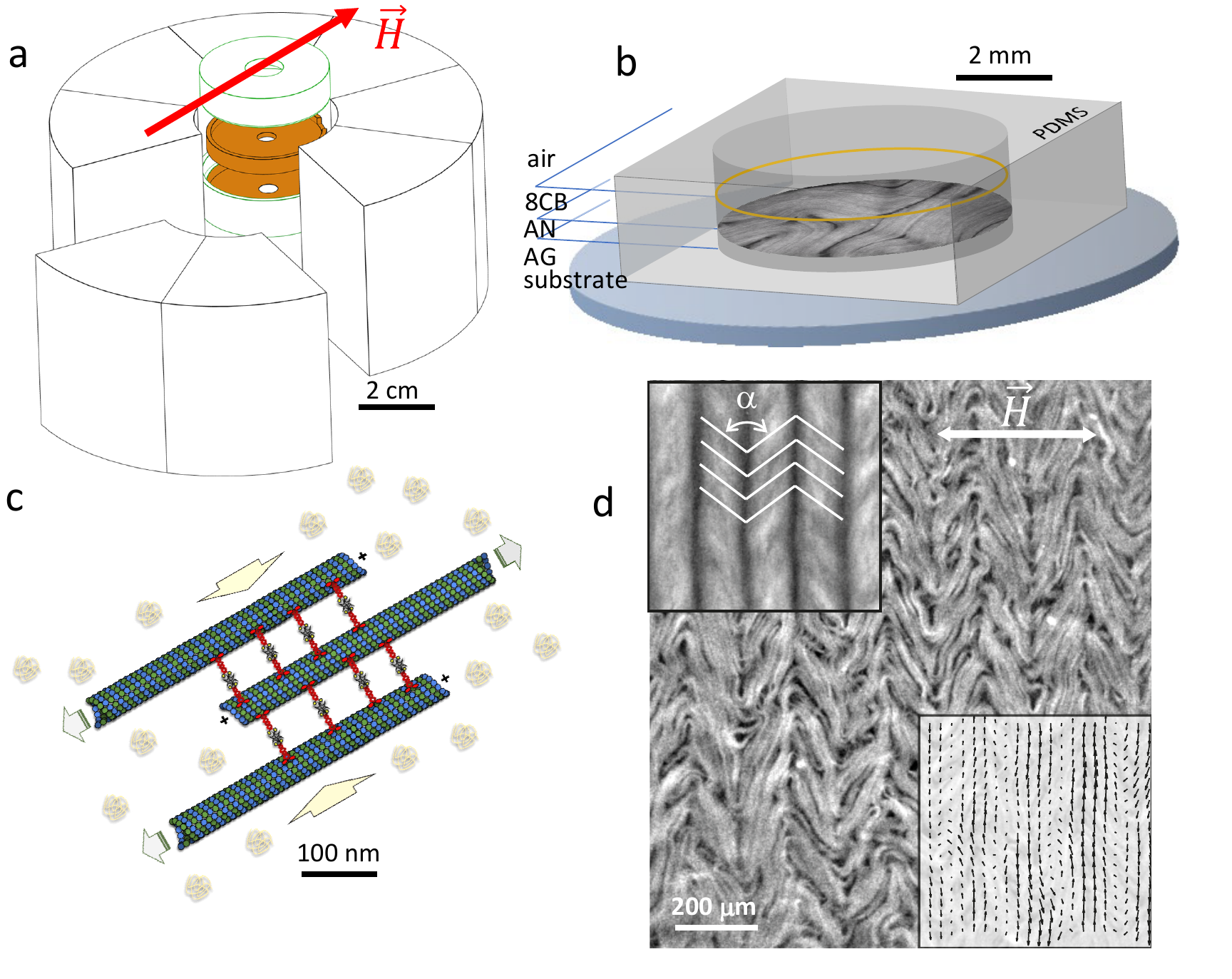}
	\caption{(a) Sketch of the setup. Two of the eight magnets that form the 1 T Halbach array are shifted to reveal the thermostatic oven that holds the sample. (b) Experimental open cell where the AN is formed at the aqueous (active gel, AG)/liquid crystal (8CB) interface. (c) Depletion forces and active cross-linking lead to the formation of the extensile active fluorescent filaments. (d) Aligned AN flows upon transition of 8CB into the SmA phase (see Movie \ref{smov:transition}). Top left inset: time-averaged fluorescence images. The active filaments organize in a chevron-like pattern (depicted by the white segments that zig-zag with an angle $\alpha$). Bottom right inset:  instantaneous velocity field. The orientation of the magnetic field, $\vec{H}$, is depicted with arrows.
}
	\label{fig:setup}
\end{figure}

Samples are enclosed by a cylindrical copper oven placed in the 25mm-wide cylindrical cavity of a 1T Halbach array formed by eight permanent magnets (Bunting Magnetics Europe Ltd.), where the magnetic field is homogeneous in a region larger than the field of view (Fig. \ref{fig:setup}a). The copper cell is equipped with heating elements (Thorlabs HT19R) and a PT100 temperature sensor (Thorlabs TH100PT), both driven by a RKC-HA400 controller that ensures a temperature stability of $0.01\,^{\circ}$C. The inner windows of the copper oven are sealed with sapphire plates to optimize thermal contact with the sample and to minimize in-plane temperature gradients across the aperture, while the outer windows are closed with glass plates for optimal thermal insulation. The active nematic layer forms at the interface between the aqueous subphase and the thermotropic liquid crystal octyl-cyanobiphenyl (8CB).
The liquids are placed inside a 6 mm-wide circular pool made from a polydimethil siloxane (PDMS) elastomer that is glued to a coverslip glass, which has been previously coated with a repulsive polyacrylamide brush \cite{Guillamat16} (Fig. \ref{fig:setup}b).
A small volume ($1\,\mu$L) of the kinesin/tubulin active mixture is placed on top of the hydrophilic glass, where it quickly spreads. The active material is prepared as described previously, with the composition detailed in Table \ref{table:Conc}.
The aqueous active mixture is readily covered by $35\,\mu$L of 8CB, which forms a layer $\simeq 1\,$mm thick on top of the much thinner active layer.
The presence in the aqueous phase of the surfactant Pluronic-127 to protect the proteins from the oil interface also favors the alignment of mesogen molecules parallel to the oil/aqueous interface, while remaining perpendicular to the oil/air interface. A combination of active flows and depletion forces (Fig. \ref{fig:setup}c) lead to the accumulation of the active filaments on the water/oil interface, forming the quasi-two-dimensional AN film that will be the object of our study \cite{Doost18}. Labeling of a fraction of the tubulin molecules allows observation of the filamentous material by means of fluorescence microscopy.

While 8CB is in the low-anisotropy nematic phase, the AN appearance does not differ measurably from the so-called active turbulent regime, characterized by chaotic flows and a proliferation of $\pm1/2$ defects moving in random directions \cite{Doost18}. Lowering the temperature below \emph{c.a.} $33.5^{\circ}$C leads to 8CB transitioning into the highly anisotropic SmA phase, in a process that is known to be continuous, or second order \cite{Oswald2005}. 8CB molecules align with the magnetic field and organize in layers perpendicular both to the field and to the interface (bookshelf configuration)\cite{Guillamat16}. While layer-sliding allows 8CB to behave as a viscous fluid in one direction, flows perpendicular to the layers are severely hindered. This determines an easy-flow direction perpendicular to the field. Active flows, hydrodynamically coupled to the anisotropic oil interface, adapt to the new environment, and become pseudo-laminar, with a velocity field that becomes perpendicular to $\vec{H}$ and develops a pattern of alternating anti-parallel flow lanes  (Fig. \ref{fig:setup}d). The latter appear dark in time-averaged fluorescence micrograph, as they concentrate the self-propelled AN defects. Simultaneously, the active filaments reorganize into chevron-like patterns, whose vertices are aligned with the flow (Fig. \ref{fig:setup}d), and whose zig-zag angle becomes smaller for higher activities (ATP concentrations).

\begin{figure}[t]
	\centering
	\includegraphics[width=\linewidth]{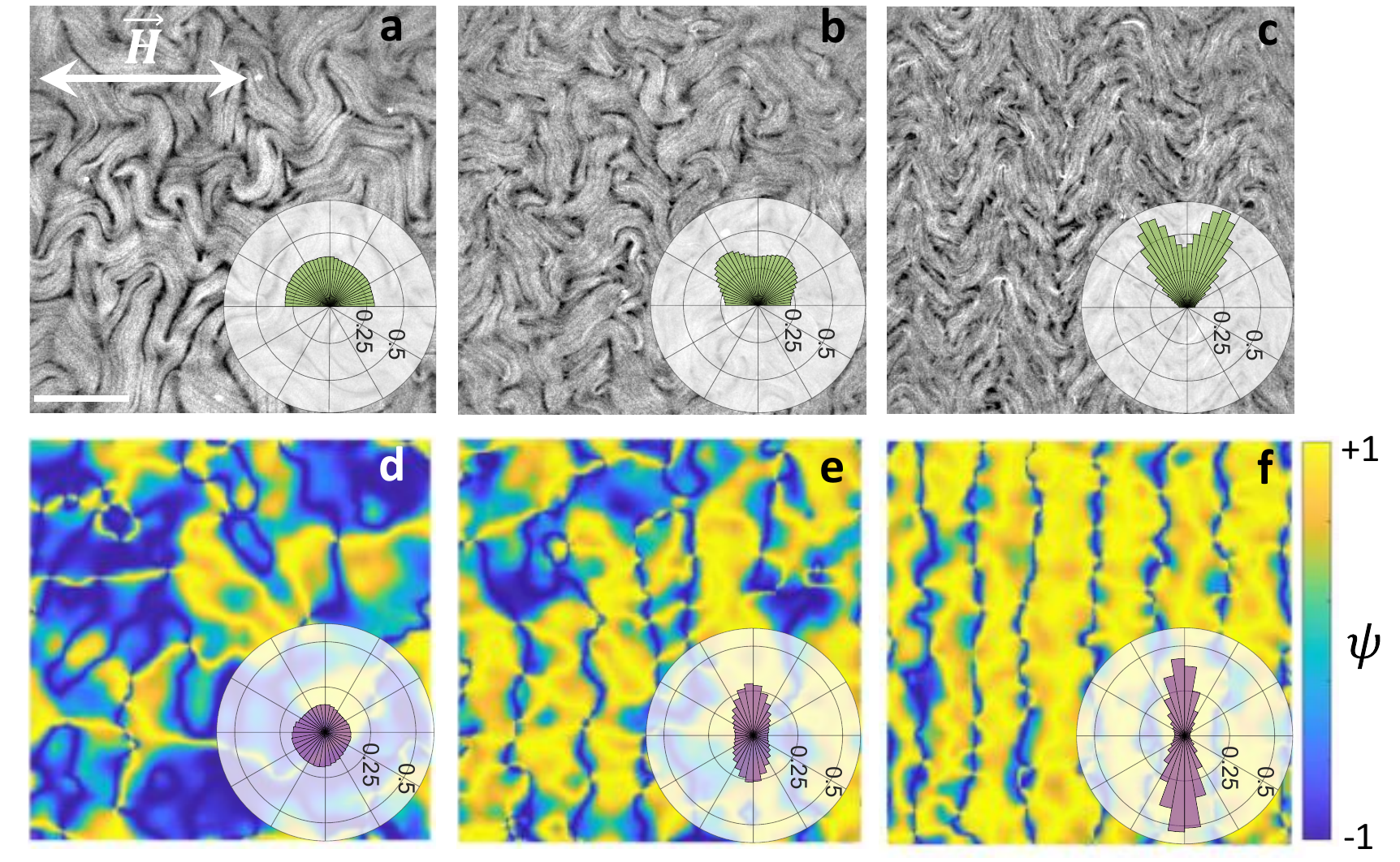}
	\caption{Alignment transition of the active nematic. The magnetic field, $\vec{H}$, is along the horizontal direction. Panels correspond to T = $33.60 ^{\circ}$C (a,d, turbulent phase), T = $33.47 ^{\circ}$C (b, e, transition), and T = $33.40 ^{\circ}$C (c, f, aligned phase). Panels a-c are instantaneous fluorescence micrographs. The insets are the distribution of director field orientations accumulated for the duration of the experiment. Panels d-f are plots of the instantaneous velocity orientation order parameter (see Movie \ref{smov:velocities}). The insets represent the distribution of velocity orientations accumulated for the duration of the experiment. Labels in all insets indicate the relative presence of different orientations.  Scale bar, $300 \mu$m.
}
	\label{fig:alignment}
\end{figure}

We then analyze the AN alignment transition as the temperature is slowly and spatially-homogeneously decreased from $33.60^{\circ}$C (8CB in nematic phase, AN in turbulent regime) down to $33.40^{\circ}$C (8CB in SmA phase, AN in aligned state).
At high temperatures, both the AN orientational and flow fields are isotropic (Fig. \ref{fig:alignment}a,d).
As the temperature is decreased, the alignment transition spans a range of about $0.10^{\circ}$C, from the first signatures of anisotropic velocity distribution at $T=33.50^{\circ}$C (Fig. \ref{sfig:alignment}) until the final steady state is reached at $T=33.40^{\circ}$C, where active flows are, on average, perpendicular to $\vec{H}$ and active filaments develop chevron-like patterns in their orientational field (Fig. \ref{fig:alignment}c,f).
Emergence of anisotropy in the latter is slightly delayed with respect to the velocity field, and is only detectable for $T<33.47^{\circ}$C (see Fig. \ref{sfig:alignment}).

\begin{figure}[t]
	\centering
	\includegraphics[width=0.9\linewidth]{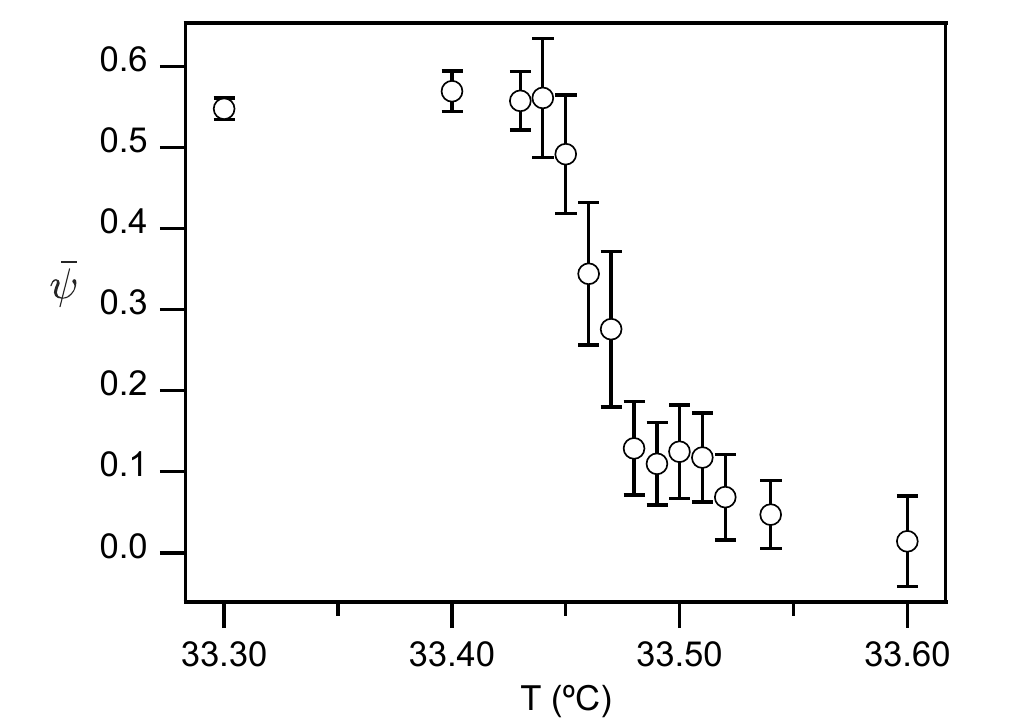}
	\caption{Evolution of the velocity orientation order parameter of the active nematic averaged in space and time, $\bar{\psi}$, as defined in the text. Temperature is changed stepwise from $33.60^{\circ}$C down to $33.30^{\circ}$C. Error bars correspond to the standard deviation of the mean order parameter at each temperature. }
	\label{fig:OP_transition}
\end{figure}

To better quantify the ordering transition we define a scalar order parameter for the orientation of the velocity field, $\psi = 2\sin^2(\theta_{\hat{v},\hat{H}})-1$, where $\theta_{\hat{v},\hat{H}}$ is the angle between the local AN velocity and the magnetic field.
While $\psi$ is randomly distributed with a mean value around 0 in the turbulent regime (Fig. \ref{fig:alignment}d), its mean value increases as we enter into the transition region (Fig. \ref{fig:alignment}e), and it is maximum in the aligned state (Fig. \ref{fig:alignment}f).
This parameter maps the flow lanes that characterize the aligned state and intermittently emerge during the transition.
To monitor the evolution from the disordered to the aligned state, we perform a statistical analysis of $\psi$ at constant temperature.
First we take a spatial average of the instantaneous distribution of $\psi$, denoted $\overline{\psi}_t$, and then compute the time average, $\overline{\psi}$, and its standard deviation over 480 s.
The result is plotted in Fig. \ref{fig:OP_transition} for different temperatures during a stepwise ramp.
We find that $\bar{\psi}$ is close to zero in the turbulent regime, consistent with a statistically isotropic distribution of orientations, and about $0.55$ in the aligned regime.

For experimental conditions where velocimetry may not be accurate enough, we have also considered an alternative definition of the order parameter, $\psi_n = 2\sin^2(\theta_{\hat{n},\hat{H}})-1$, where $\theta_{\hat{n},\hat{H}}$ now refers to the angle between the local orientational field of the AN and the magnetic field.
We observe that early stages of the transition are first picked up by $\overline{\psi}$, and that $\overline{\psi}_n$ reaches a somewhat lower steady-state value in the aligned state (see Fig. \ref{Sfig:OP_transition}).
This is expected since $\overline{\psi}_n$ depends on the angle spanned by the chevron segments, which decreases with activity.
For different experimental realisations under the same nominal conditions, the value of the order parameters can change about 5\%, likely due to small imprecisions in the concentration of ATP.

The correlation between the onset of the alignment transition in the AN layer and the N-SmA phase boundary in 8CB is better assessed by placing the sample in a temperature gradient. We created a small, in-plane gradient by unsealing the glass window that normally closes the bottom of the temperature oven.
This enables the simultaneous observation, within the field of view, of all regimes described during the alignment transition (Fig. \ref{sfig:gradient}a).
By imaging in polarization mode, we assessed the state of the 8CB layer along the gradient, revealing the precise location of the N-SmA phase boundary. We then measured $\bar{\psi_n}$ as a function of the position along the gradient (Fig. \ref{sfig:gradient}b), and found that the steepest change corresponds to the location of the N-SmA phase boundary. Consequently, in our stepwise temperature ramp experiments (Fig. \ref{fig:OP_transition} and \ref{Sfig:OP_transition}) we define $T_c$ as the temperature with the steepest change in the orientational order parameter.
We did not observe evidence of hysteresis when comparing upward and downward temperature ramps, with $T_c$ taking values that differ $\pm 0.01 ^{\circ}$C at most.
Combining data for experiments with both upward and downward ramps, and plotting a normalized $\bar{\psi_n}$ that uses the upper (aligned) and lower (turbulent) values, versus $T-T_c$, we find that all data follow the same trend (see Fig. S4).

\begin{figure}[t]
	\centering
	\includegraphics[width=\linewidth]{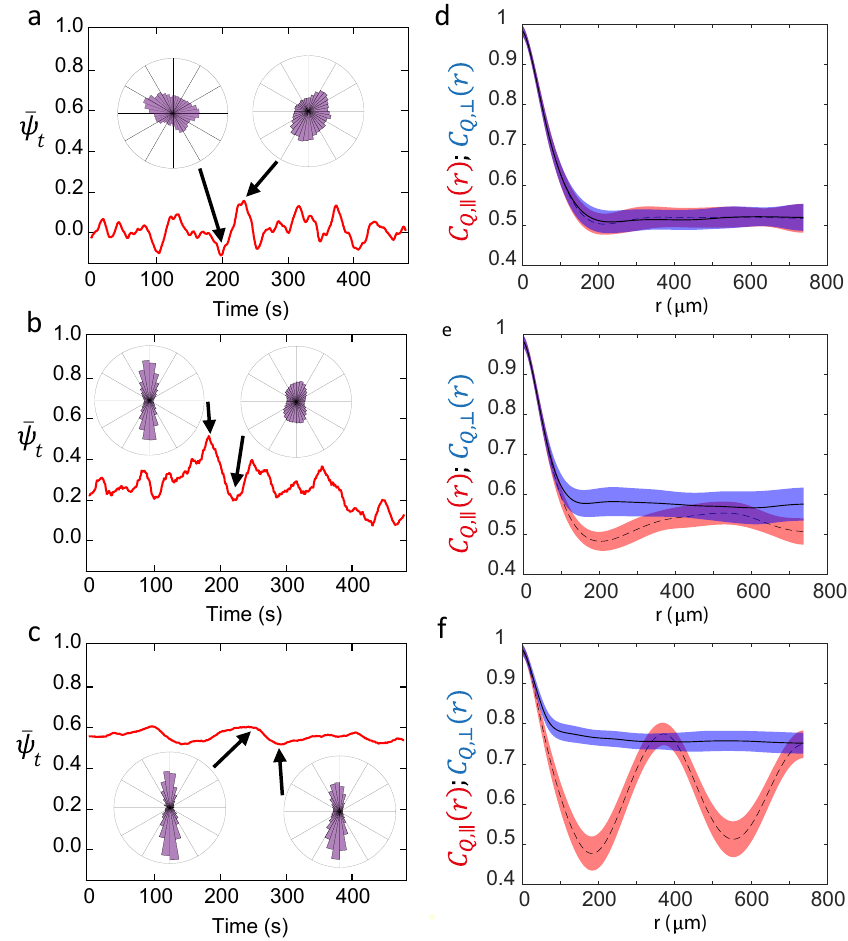}
	\caption{Onset of the alignment transition. (a-c): Temporal evolution of the spatially-averaged velocity orientation order parameter, $\bar{\psi}_t$. The insets correspond to the instantaneous distribution of velocity orientations at the times indicated by arrows. (d-f): Temporal evolution of the perpendicular to $\vec{H}$ (blue trace) and parallel to $\vec{H}$ (red trace) spatial correlation of the tensorial order parameter of the active nematic, $\mathbb{Q}$. The width of the bands are standard deviation around the time average. Data corresponds to a decreasing temperature ramp at $33.60^{\circ}$C (a,d), $33.47^{\circ}$C (b,e), and $33.40^{\circ}$C (a,d).}
	\label{fig:oscillations}
\end{figure}

As evidenced in Figs. \ref{fig:OP_transition} and \ref{Sfig:OP_transition}, the order parameters feature significant fluctuations about their mean values at constant temperature, exhibiting the highest amplitude in the transition region.  An analysis on the origin of these temporal fluctuations reveals interesting signatures of the alignment transition (Fig. \ref{fig:oscillations}).
We do this by computing the spatial correlations of the nematic tensorial order parameter, $\mathbb{Q} = 2S(\mathbf{nn}-1/2\,\mathbb{1})$ \cite{Doost18}, where $S$ is the scalar nematic order parameter and $\mathbf{n}$ is the nematic director field. These correlations quantify the extent and decay of the AN orientational order and its emergence during the alignment transition. In Fig. \ref{fig:oscillations}d-f, we measure separately the correlations in the direction parallel, $C_{Q,\parallel}(r)$ (red traces), and perpendicular, $C_{Q,\perp}(r)$ (blue traces), to the applied magnetic field.

In the turbulent regime (Fig. \ref{fig:oscillations}a), thick AN bundles adopt random orientations but with a persistence length in excess of $100\,\mu$m (Fig. \ref{fig:alignment}a). The finite size of the field of view (around 10 times the persistence length), results in temporal oscillations of $\psi_t$, rather than the steady state value $\psi_t=0$ that should correspond to a spatially-isotropic state.
The distribution of velocity orientations is quite isotropic in this regime, even when $\bar{\psi}_t$ fluctuates towards higher values (see inset in Fig. \ref{fig:oscillations}a).  The $\mathbb{Q}$-correlation curves overlap, consistently with the absence of anisotropy, and dropping their magnitude to around 0.5 at a distance corresponding to the coherence length of the unconstrained AN bundles (Fig. \ref{fig:oscillations}d).

In the aligned regime (Fig. \ref{fig:alignment}c and \ref{fig:oscillations}c), there are also oscillations in $\bar{\psi}_t$, but these occur at higher mean values and their origin is more subtle than in the turbulent regime. Thick AN bundles self-organize into the {anti-parallel flow pattern described above (Fig. \ref{fig:setup}d). The aligned bundles accumulate active stress that is released periodically in the form of bursts of bend instabilities perpendicular to the easy-flow direction that will span a certain fraction of the field of view \cite{Guillamat16}. These localized instabilities lower the instantaneous value of $\bar{\psi}_t$, which will gradually recover once the active filaments reorganize again along the easy-flow direction. In spite of these oscillations, active flows are clearly anisotropic at all times in the aligned regime (see inset in Fig. \ref{fig:oscillations}c).
We emphasize that we are using a low ATP concentration to minimize the magnitude of these oscillations (see Movie \ref{smov:transinst} for an example of the clear oscillating dynamics seen in the aligned state at ten times higher ATP concentration). In this regime, the two $\mathbb{Q}$-correlation curves are clearly separated. $C_{Q,\perp}(r)$ drops to a steady-state value of around 0.78 at a distance equal to the coherence of the AN bundles within the aligned chevrons.
Conversely, $C_{Q,\parallel}(r)$ has an oscillating behavior (Fig. \ref{fig:oscillations}f), with its maxima occurring at distances multiple of the chevron width, and its minima minima at odd multiples of the chevron half width.

With the chosen experimental conditions, oscillations in the orientational order parameters have the highest amplitude during the transition regime (Fig. \ref{fig:oscillations}b, Fig. \ref{sfig:oscillations}, and Movie \ref{smov:intermittency}), whose onset is evidenced by the emergence of spatial anisotropy upon sample cooling (Fig. \ref{fig:alignment}).
Similarly, the onset of the alignment transition can be clearly detected in Fig. \ref{fig:oscillations}e, where the traces of $C_{Q,\perp}(r)$ and $C_{Q,\parallel}(r)$ are already distinguishable and do not overlap like in the turbulent regime.
The trace of $\bar{\psi}_t$ contains large amplitude oscillations, whose minima coincide with episodes of low anisotropy in the velocity field, while the maxima arise when velocity anisotropy is high (see insets in Fig. \ref{fig:oscillations}b). We have ruled out noise from the temperature controller as the source of these oscillations by comparing the temporal fluctuations of temperature with the oscillations in $\bar{\psi}_t$ (see Fig. \ref{sfig:Tfluctuations}).

The reported oscillating behavior in the transition region evidences both intermittent dynamics of the order parameter and spatial coexistence of turbulent and aligned regions, a signature that this transition is, in fact, discontinuous or first order. The abrupt, yet continuous change in $\bar{\psi}$ reported in Fig. \ref{fig:OP_transition} could, nevertheless, be compatible with a continuous transition between the aligned and disordered regimes. A continuous transition, however, would be inconsistent with the alternation between high anisotropy and low anisotropy configurations that coincide with large-amplitude oscillations in $\bar{\psi}_t$ in the transition region (Fig. \ref{fig:oscillations}b and Movie \ref{smov:intermittency}). Indeed, during the alignment transition, localized regions of random size become ordered and remain in that state for a period of time. When their span is comparable to the field of view, $\bar{\psi}_t$ attains a maximum. While identifying these aligned domains may be difficult from the fluorescence micrographs, they are more easily revealed when considering the velocity field (see Fig. \ref{sfig:coexistence} and Movie \ref{smov:intermittency}). Conversely, when most of the field of view is in the disordered state, $\bar{\psi}_t$ attains a local minimum. Finally, when small ordered and disordered domains coexist, they average out in the field of view, and $\bar{\psi}_t$ features only small amplitude fluctuations about its mean value, and the complex nature of this dynamic transition is not readily apparent. In Fig. \ref{sfig:coexistence}, we have identified some instances of the spatial coexistence between aligned and disordered regions during the alignment transition. In those examples, we indicate the contrasted instantaneous value of $\bar{\psi}_t$ measured inside and outside the aligned regions.

The fact that this dynamic transition is discontinuous, yet it exhibits a continuous change in the order parameter, could be interpreted as a smoothening due to finite size effects. This is, in fact, not unusual in active matter, where the number of constituent building blocks is typically small, always far from atomic or molecular systems where phase transitions are usually studied. Moreover, as reported above, the observed temperature span of the dynamic transition is around $0.1 ^{\circ}$, centered at $T_{N-SmA}$, the N-SmA transition temperature of 8CB.
It is known that the nematic phase of thermotropic LCs features pre-transitional effects where regions with smectic-like order, or cybotactic groups, emerge close to $T_{N-SmA}$ \cite{oswald2006}. This is likely to have an impact on the rheology of the mesophase, and could justify that AN alignment starts at temperatures slightly above $T_{N-SmA}$.
In addition, coupling with the flowing AN results in energy injection into the LC at a broad range of length scales \cite{Alert19}. This may trigger anomalously high fluctuations in the SmA phase close to the transition temperature, and could justify that the AN alignment transition is not completed until the temperature is slightly below $T_{N-SmA}$.

In the reported experiments, both the passive (8CB) and active (AN) subsystems change their inherent symmetries at the crossover temperature, but in a very different way and affecting very different length scales.
While 8CB nematogens gain spatial correlations in transitioning from the positionally-disordered nematic arrangement to the layered configuration of the smectic A phase, the concurrent AN flows change from the isotropic turbulent to an aligned, quasi-laminar regime.
Importantly, the order of either phase transition is different; it is continuous in the passive phase and discontinuous in the active phase.
Future work to unravel the origin of this fascinating difference calls for new developments in the statistical physics of active matter.

The authors are indebted to the Brandeis University MRSEC Biosynthesis facility for providing the tubulin. We thank M. Pons, A. LeRoux, and G. Iruela (Universitat de Barcelona) for their assistance in the expression of motor proteins. O. B., J.I.-M., and F.S. acknowledge funding from MICINN/AEI/10.13039/501100011033 (Grant No. PID2019-108842GB-C22).  J.N. and A.F-N. acknowledge funding from  MCINN/AEI/10.13039/501100011033/FEDER, UE (Grant No. PID2021-122369NB-100). Brandeis University MRSEC Biosynthesis facility is supported by NSF MRSEC 2011846. The authors acknowledge helpful discussions with H. Chat\'e.


\bibliography{alignment}

\begin{thebibliography}{22}%
\makeatletter
\providecommand \@ifxundefined [1]{%
 \@ifx{#1\undefined}
}%
\providecommand \@ifnum [1]{%
 \ifnum #1\expandafter \@firstoftwo
 \else \expandafter \@secondoftwo
 \fi
}%
\providecommand \@ifx [1]{%
 \ifx #1\expandafter \@firstoftwo
 \else \expandafter \@secondoftwo
 \fi
}%
\providecommand \natexlab [1]{#1}%
\providecommand \enquote  [1]{``#1''}%
\providecommand \bibnamefont  [1]{#1}%
\providecommand \bibfnamefont [1]{#1}%
\providecommand \citenamefont [1]{#1}%
\providecommand \href@noop [0]{\@secondoftwo}%
\providecommand \href [0]{\begingroup \@sanitize@url \@href}%
\providecommand \@href[1]{\@@startlink{#1}\@@href}%
\providecommand \@@href[1]{\endgroup#1\@@endlink}%
\providecommand \@sanitize@url [0]{\catcode `\\12\catcode `\$12\catcode
  `\&12\catcode `\#12\catcode `\^12\catcode `\_12\catcode `\%12\relax}%
\providecommand \@@startlink[1]{}%
\providecommand \@@endlink[0]{}%
\providecommand \url  [0]{\begingroup\@sanitize@url \@url }%
\providecommand \@url [1]{\endgroup\@href {#1}{\urlprefix }}%
\providecommand \urlprefix  [0]{URL }%
\providecommand \Eprint [0]{\href }%
\providecommand \doibase [0]{https://doi.org/}%
\providecommand \selectlanguage [0]{\@gobble}%
\providecommand \bibinfo  [0]{\@secondoftwo}%
\providecommand \bibfield  [0]{\@secondoftwo}%
\providecommand \translation [1]{[#1]}%
\providecommand \BibitemOpen [0]{}%
\providecommand \bibitemStop [0]{}%
\providecommand \bibitemNoStop [0]{.\EOS\space}%
\providecommand \EOS [0]{\spacefactor3000\relax}%
\providecommand \BibitemShut  [1]{\csname bibitem#1\endcsname}%
\let\auto@bib@innerbib\@empty
\bibitem [{\citenamefont {Ramaswamy}(2010)}]{Ramaswamy10}%
  \BibitemOpen
  \bibfield  {author} {\bibinfo {author} {\bibfnamefont {S.}~\bibnamefont
  {Ramaswamy}},\ }\bibfield  {title} {\bibinfo {title} {{The Mechanics and
  Statistics of Active Matter}},\ }\href@noop {} {\bibfield  {journal}
  {\bibinfo  {journal} {Annu. Rev. Condens.Matter Phys.}\ }\textbf {\bibinfo
  {volume} {1}},\ \bibinfo {pages} {323} (\bibinfo {year} {2010})}\BibitemShut
  {NoStop}%
\bibitem [{\citenamefont {Marchetti}\ \emph {et~al.}(2013)\citenamefont
  {Marchetti}, \citenamefont {Joanny}, \citenamefont {Ramaswamy}, \citenamefont
  {Liverpool}, \citenamefont {Prost}, \citenamefont {Rao},\ and\ \citenamefont
  {Simha}}]{marchetti13}%
  \BibitemOpen
  \bibfield  {author} {\bibinfo {author} {\bibfnamefont {M.~C.}\ \bibnamefont
  {Marchetti}}, \bibinfo {author} {\bibfnamefont {J.~F.}\ \bibnamefont
  {Joanny}}, \bibinfo {author} {\bibfnamefont {S.}~\bibnamefont {Ramaswamy}},
  \bibinfo {author} {\bibfnamefont {T.~B.}\ \bibnamefont {Liverpool}}, \bibinfo
  {author} {\bibfnamefont {J.}~\bibnamefont {Prost}}, \bibinfo {author}
  {\bibfnamefont {M.}~\bibnamefont {Rao}},\ and\ \bibinfo {author}
  {\bibfnamefont {R.~A.}\ \bibnamefont {Simha}},\ }\bibfield  {title} {\bibinfo
  {title} {Hydrodynamics of soft active matter},\ }\href@noop {} {\bibfield
  {journal} {\bibinfo  {journal} {Reviews of Modern Physics}\ }\textbf
  {\bibinfo {volume} {85}},\ \bibinfo {pages} {1143} (\bibinfo {year}
  {2013})}\BibitemShut {NoStop}%
\bibitem [{\citenamefont {Sagu\'es}(2022)}]{sagues22}%
  \BibitemOpen
  \bibfield  {author} {\bibinfo {author} {\bibfnamefont {F.}~\bibnamefont
  {Sagu\'es}},\ }\href@noop {} {\emph {\bibinfo {title} {Colloidal Active
  Matter: Concepts, Experimental Realizations, and Models (1st ed.).}}}\
  (\bibinfo {year} {2022})\BibitemShut {NoStop}%
\bibitem [{\citenamefont {Cates}\ and\ \citenamefont
  {Tailleur}(2015)}]{cates15}%
  \BibitemOpen
  \bibfield  {author} {\bibinfo {author} {\bibfnamefont {M.~E.}\ \bibnamefont
  {Cates}}\ and\ \bibinfo {author} {\bibfnamefont {J.}~\bibnamefont
  {Tailleur}},\ }\bibfield  {title} {\bibinfo {title} {Motility-induced phase
  separation},\ }\href
  {https://doi.org/10.1146/annurev-conmatphys-031214-014710} {\bibfield
  {journal} {\bibinfo  {journal} {Ann. Rev. Cond. Matt. Phys.}\ }\textbf
  {\bibinfo {volume} {6}},\ \bibinfo {pages} {219} (\bibinfo {year}
  {2015})}\BibitemShut {NoStop}%
\bibitem [{\citenamefont {van~der Linden}\ \emph {et~al.}(2019)\citenamefont
  {van~der Linden}, \citenamefont {Alexander}, \citenamefont {Aarts},\ and\
  \citenamefont {Dauchot}}]{vanderlinden19}%
  \BibitemOpen
  \bibfield  {author} {\bibinfo {author} {\bibfnamefont {M.~N.}\ \bibnamefont
  {van~der Linden}}, \bibinfo {author} {\bibfnamefont {L.~C.}\ \bibnamefont
  {Alexander}}, \bibinfo {author} {\bibfnamefont {D.}~\bibnamefont {Aarts}},\
  and\ \bibinfo {author} {\bibfnamefont {O.}~\bibnamefont {Dauchot}},\
  }\bibfield  {title} {\bibinfo {title} {Interrupted motility induced phase
  separation in aligning active colloids},\ }\href@noop {} {\bibfield
  {journal} {\bibinfo  {journal} {Phys Rev Lett}\ }\textbf {\bibinfo {volume}
  {123}},\ \bibinfo {pages} {098001} (\bibinfo {year} {2019})}\BibitemShut
  {NoStop}%
\bibitem [{\citenamefont {Geyer}\ \emph {et~al.}(2019)\citenamefont {Geyer},
  \citenamefont {Martin}, \citenamefont {Tailleur},\ and\ \citenamefont
  {Bartolo}}]{geyer19}%
  \BibitemOpen
  \bibfield  {author} {\bibinfo {author} {\bibfnamefont {D.}~\bibnamefont
  {Geyer}}, \bibinfo {author} {\bibfnamefont {D.}~\bibnamefont {Martin}},
  \bibinfo {author} {\bibfnamefont {J.}~\bibnamefont {Tailleur}},\ and\
  \bibinfo {author} {\bibfnamefont {D.}~\bibnamefont {Bartolo}},\ }\bibfield
  {title} {\bibinfo {title} {Freezing a flock: Motility-induced phase
  separation in polar active liquids},\ }\href@noop {} {\bibfield  {journal}
  {\bibinfo  {journal} {Physical Review X}\ }\textbf {\bibinfo {volume} {9}},\
  \bibinfo {pages} {031043} (\bibinfo {year} {2019})}\BibitemShut {NoStop}%
\bibitem [{\citenamefont {Chat\'{e}}(2020)}]{chate20}%
  \BibitemOpen
  \bibfield  {author} {\bibinfo {author} {\bibfnamefont {H.}~\bibnamefont
  {Chat\'{e}}},\ }\bibfield  {title} {\bibinfo {title} {Dry aligning dilute
  active matter},\ }\href@noop {} {\bibfield  {journal} {\bibinfo  {journal}
  {Ann. Rev. Cond. Matt. Phys.}\ }\textbf {\bibinfo {volume} {11}},\ \bibinfo
  {pages} {189} (\bibinfo {year} {2020})}\BibitemShut {NoStop}%
\bibitem [{\citenamefont {Cates}\ \emph {et~al.}(2010)\citenamefont {Cates},
  \citenamefont {Marenduzzo}, \citenamefont {Pagonabarraga},\ and\
  \citenamefont {Tailleur}}]{cates10}%
  \BibitemOpen
  \bibfield  {author} {\bibinfo {author} {\bibfnamefont {M.~E.}\ \bibnamefont
  {Cates}}, \bibinfo {author} {\bibfnamefont {D.}~\bibnamefont {Marenduzzo}},
  \bibinfo {author} {\bibfnamefont {I.}~\bibnamefont {Pagonabarraga}},\ and\
  \bibinfo {author} {\bibfnamefont {J.}~\bibnamefont {Tailleur}},\ }\bibfield
  {title} {\bibinfo {title} {Arrested phase separation in reproducing bacteria
  creates a generic route to pattern formation},\ }\href@noop {} {\bibfield
  {journal} {\bibinfo  {journal} {Proc Natl Acad Sci U S A}\ }\textbf {\bibinfo
  {volume} {107}},\ \bibinfo {pages} {11715} (\bibinfo {year}
  {2010})}\BibitemShut {NoStop}%
\bibitem [{\citenamefont {Schwarz-Linek}\ \emph {et~al.}(2012)\citenamefont
  {Schwarz-Linek}, \citenamefont {Valeriani}, \citenamefont {Cacciuto},
  \citenamefont {Cates}, \citenamefont {Marenduzzo}, \citenamefont {Morozov},\
  and\ \citenamefont {Poon}}]{schwarz12}%
  \BibitemOpen
  \bibfield  {author} {\bibinfo {author} {\bibfnamefont {J.}~\bibnamefont
  {Schwarz-Linek}}, \bibinfo {author} {\bibfnamefont {C.}~\bibnamefont
  {Valeriani}}, \bibinfo {author} {\bibfnamefont {A.}~\bibnamefont {Cacciuto}},
  \bibinfo {author} {\bibfnamefont {M.~E.}\ \bibnamefont {Cates}}, \bibinfo
  {author} {\bibfnamefont {D.}~\bibnamefont {Marenduzzo}}, \bibinfo {author}
  {\bibfnamefont {A.~N.}\ \bibnamefont {Morozov}},\ and\ \bibinfo {author}
  {\bibfnamefont {W.~C.}\ \bibnamefont {Poon}},\ }\bibfield  {title} {\bibinfo
  {title} {Phase separation and rotor self-assembly in active particle
  suspensions},\ }\href@noop {} {\bibfield  {journal} {\bibinfo  {journal}
  {Proc Natl Acad Sci U S A}\ }\textbf {\bibinfo {volume} {109}},\ \bibinfo
  {pages} {4052} (\bibinfo {year} {2012})}\BibitemShut {NoStop}%
\bibitem [{\citenamefont {Blow}\ \emph {et~al.}(2014)\citenamefont {Blow},
  \citenamefont {Thampi},\ and\ \citenamefont {Yeomans}}]{blow14}%
  \BibitemOpen
  \bibfield  {author} {\bibinfo {author} {\bibfnamefont {M.~L.}\ \bibnamefont
  {Blow}}, \bibinfo {author} {\bibfnamefont {S.~P.}\ \bibnamefont {Thampi}},\
  and\ \bibinfo {author} {\bibfnamefont {J.~M.}\ \bibnamefont {Yeomans}},\
  }\bibfield  {title} {\bibinfo {title} {Biphasic, lyotropic, active
  nematics},\ }\href@noop {} {\bibfield  {journal} {\bibinfo  {journal} {Phys.
  Rev. Lett.}\ }\textbf {\bibinfo {volume} {113}},\ \bibinfo {pages} {248303}
  (\bibinfo {year} {2014})}\BibitemShut {NoStop}%
\bibitem [{\citenamefont {Soni}\ \emph {et~al.}(2019)\citenamefont {Soni},
  \citenamefont {Bililign}, \citenamefont {Magkiriadou}, \citenamefont
  {Sacanna}, \citenamefont {Bartolo}, \citenamefont {Shelley},\ and\
  \citenamefont {Irvine}}]{soni19}%
  \BibitemOpen
  \bibfield  {author} {\bibinfo {author} {\bibfnamefont {V.}~\bibnamefont
  {Soni}}, \bibinfo {author} {\bibfnamefont {E.~S.}\ \bibnamefont {Bililign}},
  \bibinfo {author} {\bibfnamefont {S.}~\bibnamefont {Magkiriadou}}, \bibinfo
  {author} {\bibfnamefont {S.}~\bibnamefont {Sacanna}}, \bibinfo {author}
  {\bibfnamefont {D.}~\bibnamefont {Bartolo}}, \bibinfo {author} {\bibfnamefont
  {M.~J.}\ \bibnamefont {Shelley}},\ and\ \bibinfo {author} {\bibfnamefont
  {W.~T.~M.}\ \bibnamefont {Irvine}},\ }\bibfield  {title} {\bibinfo {title}
  {The odd free surface flows of a colloidal chiral fluid},\ }\href@noop {}
  {\bibfield  {journal} {\bibinfo  {journal} {Nature Physics}\ }\textbf
  {\bibinfo {volume} {15}},\ \bibinfo {pages} {1188} (\bibinfo {year}
  {2019})}\BibitemShut {NoStop}%
\bibitem [{\citenamefont {Adkins}\ \emph {et~al.}(2022)\citenamefont {Adkins},
  \citenamefont {Kolvin}, \citenamefont {You}, \citenamefont {Witthaus},
  \citenamefont {Marchetti},\ and\ \citenamefont {Dogic}}]{adkins22}%
  \BibitemOpen
  \bibfield  {author} {\bibinfo {author} {\bibfnamefont {R.}~\bibnamefont
  {Adkins}}, \bibinfo {author} {\bibfnamefont {I.}~\bibnamefont {Kolvin}},
  \bibinfo {author} {\bibfnamefont {Z.}~\bibnamefont {You}}, \bibinfo {author}
  {\bibfnamefont {S.}~\bibnamefont {Witthaus}}, \bibinfo {author}
  {\bibfnamefont {M.~C.}\ \bibnamefont {Marchetti}},\ and\ \bibinfo {author}
  {\bibfnamefont {Z.}~\bibnamefont {Dogic}},\ }\bibfield  {title} {\bibinfo
  {title} {Dynamics of active liquid interfaces},\ }\href@noop {} {\bibfield
  {journal} {\bibinfo  {journal} {Science}\ }\textbf {\bibinfo {volume}
  {377}},\ \bibinfo {pages} {768} (\bibinfo {year} {2022})}\BibitemShut
  {NoStop}%
\bibitem [{\citenamefont {Sanchez}\ \emph {et~al.}(2012)\citenamefont
  {Sanchez}, \citenamefont {Chen}, \citenamefont {DeCamp}, \citenamefont
  {Heymann},\ and\ \citenamefont {Dogic}}]{sanchez12}%
  \BibitemOpen
  \bibfield  {author} {\bibinfo {author} {\bibfnamefont {T.}~\bibnamefont
  {Sanchez}}, \bibinfo {author} {\bibfnamefont {D.~T.}\ \bibnamefont {Chen}},
  \bibinfo {author} {\bibfnamefont {S.~J.}\ \bibnamefont {DeCamp}}, \bibinfo
  {author} {\bibfnamefont {M.}~\bibnamefont {Heymann}},\ and\ \bibinfo {author}
  {\bibfnamefont {Z.}~\bibnamefont {Dogic}},\ }\bibfield  {title} {\bibinfo
  {title} {Spontaneous motion in hierarchically assembled active matter},\
  }\href@noop {} {\bibfield  {journal} {\bibinfo  {journal} {Nature}\ }\textbf
  {\bibinfo {volume} {491}},\ \bibinfo {pages} {431} (\bibinfo {year}
  {2012})}\BibitemShut {NoStop}%
\bibitem [{\citenamefont {Doostmohammadi}\ \emph {et~al.}(2018)\citenamefont
  {Doostmohammadi}, \citenamefont {Ign{\'{e}}s-Mullol}, \citenamefont
  {Yeomans},\ and\ \citenamefont {Sagu{\'{e}}s}}]{Doost18}%
  \BibitemOpen
  \bibfield  {author} {\bibinfo {author} {\bibfnamefont {A.}~\bibnamefont
  {Doostmohammadi}}, \bibinfo {author} {\bibfnamefont {J.}~\bibnamefont
  {Ign{\'{e}}s-Mullol}}, \bibinfo {author} {\bibfnamefont {J.~M.}\ \bibnamefont
  {Yeomans}},\ and\ \bibinfo {author} {\bibfnamefont {F.}~\bibnamefont
  {Sagu{\'{e}}s}},\ }\bibfield  {title} {\bibinfo {title} {{Active nematics}},\
  }\href {https://doi.org/10.1038/s41467-018-05666-8} {\bibfield  {journal}
  {\bibinfo  {journal} {Nat. Commun.}\ }\textbf {\bibinfo {volume} {9}},\
  \bibinfo {pages} {3246} (\bibinfo {year} {2018})}\BibitemShut {NoStop}%
\bibitem [{\citenamefont {Giomi}(2015)}]{giomi15}%
  \BibitemOpen
  \bibfield  {author} {\bibinfo {author} {\bibfnamefont {L.}~\bibnamefont
  {Giomi}},\ }\bibfield  {title} {\bibinfo {title} {Geometry and topology of
  turbulence in active nematics},\ }\href
  {https://doi.org/10.1103/PhysRevX.5.031003} {\bibfield  {journal} {\bibinfo
  {journal} {Phys. Rev. X}\ }\textbf {\bibinfo {volume} {5}},\ \bibinfo {pages}
  {031003} (\bibinfo {year} {2015})}\BibitemShut {NoStop}%
\bibitem [{\citenamefont {Urzay}\ \emph {et~al.}(2017)\citenamefont {Urzay},
  \citenamefont {Doostmohammadi},\ and\ \citenamefont {Yeomans}}]{urzay17}%
  \BibitemOpen
  \bibfield  {author} {\bibinfo {author} {\bibfnamefont {J.}~\bibnamefont
  {Urzay}}, \bibinfo {author} {\bibfnamefont {A.}~\bibnamefont
  {Doostmohammadi}},\ and\ \bibinfo {author} {\bibfnamefont {J.~M.}\
  \bibnamefont {Yeomans}},\ }\bibfield  {title} {\bibinfo {title} {{Multi-scale
  statistics of turbulence motorized by active matter}},\ }\href
  {https://doi.org/10.1017/jfm.2017.311} {\bibfield  {journal} {\bibinfo
  {journal} {J. Fluid Mech.}\ }\textbf {\bibinfo {volume} {822}},\ \bibinfo
  {pages} {762} (\bibinfo {year} {2017})}\BibitemShut {NoStop}%
\bibitem [{\citenamefont {Mart{\'{i}}nez-Prat}\ \emph
  {et~al.}(2019)\citenamefont {Mart{\'{i}}nez-Prat}, \citenamefont
  {Ign{\'{e}}s-Mullol}, \citenamefont {Casademunt},\ and\ \citenamefont
  {Sagu{\'{e}}s}}]{martinez19}%
  \BibitemOpen
  \bibfield  {author} {\bibinfo {author} {\bibfnamefont {B.}~\bibnamefont
  {Mart{\'{i}}nez-Prat}}, \bibinfo {author} {\bibfnamefont {J.}~\bibnamefont
  {Ign{\'{e}}s-Mullol}}, \bibinfo {author} {\bibfnamefont {J.}~\bibnamefont
  {Casademunt}},\ and\ \bibinfo {author} {\bibfnamefont {F.}~\bibnamefont
  {Sagu{\'{e}}s}},\ }\bibfield  {title} {\bibinfo {title} {{Selection mechanism
  at the onset of active turbulence}},\ }\href
  {https://doi.org/10.1038/s41567-018-0411-6} {\bibfield  {journal} {\bibinfo
  {journal} {Nature Physics}\ }\textbf {\bibinfo {volume} {15}},\ \bibinfo
  {pages} {362} (\bibinfo {year} {2019})}\BibitemShut {NoStop}%
\bibitem [{\citenamefont {Oswald}\ and\ \citenamefont
  {Pieranski}(2006)}]{oswald2006}%
  \BibitemOpen
  \bibfield  {author} {\bibinfo {author} {\bibfnamefont {P.}~\bibnamefont
  {Oswald}}\ and\ \bibinfo {author} {\bibfnamefont {P.}~\bibnamefont
  {Pieranski}},\ }\href@noop {} {\emph {\bibinfo {title} {Smectic and columnar
  liquid crystals : concepts and physical properties illustrated by
  experiments}}}\ (\bibinfo  {publisher} {Taylor \& Francis},\ \bibinfo
  {address} {Boca Raton, FL},\ \bibinfo {year} {2006})\BibitemShut {NoStop}%
\bibitem [{\citenamefont {Guillamat}\ \emph {et~al.}(2016)\citenamefont
  {Guillamat}, \citenamefont {Ign\'es-Mullol},\ and\ \citenamefont
  {Sagu\'es}}]{Guillamat16}%
  \BibitemOpen
  \bibfield  {author} {\bibinfo {author} {\bibfnamefont {P.}~\bibnamefont
  {Guillamat}}, \bibinfo {author} {\bibfnamefont {J.}~\bibnamefont
  {Ign\'es-Mullol}},\ and\ \bibinfo {author} {\bibfnamefont {F.}~\bibnamefont
  {Sagu\'es}},\ }\bibfield  {title} {\bibinfo {title} {Control of active liquid
  crystals with a magnetic field},\ }\href@noop {} {\bibfield  {journal}
  {\bibinfo  {journal} {Proc. Natl. Acad. Sci.}\ }\textbf {\bibinfo {volume}
  {113}},\ \bibinfo {pages} {5498} (\bibinfo {year} {2016})}\BibitemShut
  {NoStop}%
\bibitem [{\citenamefont {Guillamat}\ \emph {et~al.}(2017)\citenamefont
  {Guillamat}, \citenamefont {Ign\'es-Mullol},\ and\ \citenamefont
  {Sagu\'es}}]{Guillamat17}%
  \BibitemOpen
  \bibfield  {author} {\bibinfo {author} {\bibfnamefont {P.}~\bibnamefont
  {Guillamat}}, \bibinfo {author} {\bibfnamefont {J.}~\bibnamefont
  {Ign\'es-Mullol}},\ and\ \bibinfo {author} {\bibfnamefont {F.}~\bibnamefont
  {Sagu\'es}},\ }\bibfield  {title} {\bibinfo {title} {Taming active turbulence
  with patterned soft interfaces},\ }\href@noop {} {\bibfield  {journal}
  {\bibinfo  {journal} {Nat. Commun.}\ }\textbf {\bibinfo {volume} {8}},\
  \bibinfo {pages} {564} (\bibinfo {year} {2017})}\BibitemShut {NoStop}%
\bibitem [{\citenamefont {Oswald}\ and\ \citenamefont
  {Pieranski}(2005)}]{Oswald2005}%
  \BibitemOpen
  \bibfield  {author} {\bibinfo {author} {\bibfnamefont {P.}~\bibnamefont
  {Oswald}}\ and\ \bibinfo {author} {\bibfnamefont {P.}~\bibnamefont
  {Pieranski}},\ }\href@noop {} {\emph {\bibinfo {title} {Nematic and
  cholesteric liquid crystals : concepts and physical properties illustrated by
  experiments}}}\ (\bibinfo  {publisher} {Taylor \& Francis},\ \bibinfo {year}
  {2005})\BibitemShut {NoStop}%
\bibitem [{\citenamefont {Alert}\ \emph {et~al.}(2020)\citenamefont {Alert},
  \citenamefont {Joanny},\ and\ \citenamefont {Casademunt}}]{Alert19}%
  \BibitemOpen
  \bibfield  {author} {\bibinfo {author} {\bibfnamefont {R.}~\bibnamefont
  {Alert}}, \bibinfo {author} {\bibfnamefont {J.-F.}\ \bibnamefont {Joanny}},\
  and\ \bibinfo {author} {\bibfnamefont {J.}~\bibnamefont {Casademunt}},\
  }\bibfield  {title} {\bibinfo {title} {{Universal scaling of active nematic
  turbulence}},\ }\href {https://doi.org/10.1038/s41567-020-0854-4} {\bibfield
  {journal} {\bibinfo  {journal} {Nat. Phys.}\ }\textbf {\bibinfo {volume}
  {16}},\ \bibinfo {pages} {682} (\bibinfo {year} {2020})}\BibitemShut
  {NoStop}%
\end{thebibliography}%


\clearpage

\setcounter{page}{1}
\setcounter{section}{0}

\beginsupplement

\onecolumngrid
\begin{center}
\bf{SUPPLEMENTARY MATERIAL}
\end{center}

\vspace{1cm}

\begin{center}
{\bf First order alignment transition in an interfaced active nematic}\\
\vspace{5mm}
Bantysh \sl{et al.}
\end{center}


\onecolumngrid

\vspace{1cm}

\begin{center}
{\bf Supplementary Table}
\end{center}

\begin{table}[!ht]
    \begin{tabular}{|l|l|c|l|}
    \hline
        \textbf{Compound} & \textbf{Buffer} & \textbf{Final Conc.} & \textbf{Units}  \\ \hline
        PEG (20 kDa) & M2B & 1.54 & \% w/v  \\ \hline
        PEP & M2B & 25.68 & mM  \\ \hline
        MgCl$_2$ & M2B & 3.12 & mM  \\ \hline
        ATP & M2B & 137 & $\mu$M  \\ \hline
        Pluronic-147 & M2B & 0.41 & w/v  \\ \hline
        DTT & M2B & 5.21 & mM  \\ \hline
        Streptavidin & M2B & 0.01 & mg/mL  \\ \hline
        Trolox & Phosphate & 1.93 & mM  \\ \hline
        Catalase & Phosphate & 0.04 & mg/mL  \\ \hline
        Glucose & Phosphate & 3.20 & mg/mL  \\ \hline
        Glucose Oxidase & Phosphate & 0.21 & mg/mL  \\ \hline
        PK & Original & 25.01 & u/mL  \\ \hline
        LDH & Original & 24.91 & u/mL  \\ \hline
        Kinesin & Original & 0.08 & mg/mL  \\ \hline
        Microtubules & Original & 1.85 & mg/mL  \\ \hline
    \end{tabular}
    \caption{Composition of all stock solutions (including buffer used for their preparation), and concentration of the different species in the final mixture. Acronyms used in this table are: PEG (Poly-ethylene glycol); PEP (Phosphoenol pyruvate); ATP (Adenosin triphosphate); PK (Pyruvate Kinase); LDH (Lactic Dehydrogenase); DTT (1,4-dithiothreitol). M2B buffer: 80 mM PIPES (piperazine-N,N'-bis(2-ethanesulfonic acid)) pH 6.8, 2 mM MgCl$_2$ 1 mM EGTA (egtazic acid). Phosphate buffer: 20 mM Phosphate buffer (6.68 mM KH$_2$PO$_4$, 12.32 mM K$_2$HPO$_4$) pH 7.2;  Original: species is obtained already dissolved in its custom buffer. Note that the concentration of ATP is ten times lower than usual in this preparation. }
    \label{table:Conc}
\end{table}

\cleardoublepage
\begin{center}
{\bf Supplementary Figures}
\end{center}

\begin{figure}[h]
	\centering
	\includegraphics[width=0.57\linewidth]{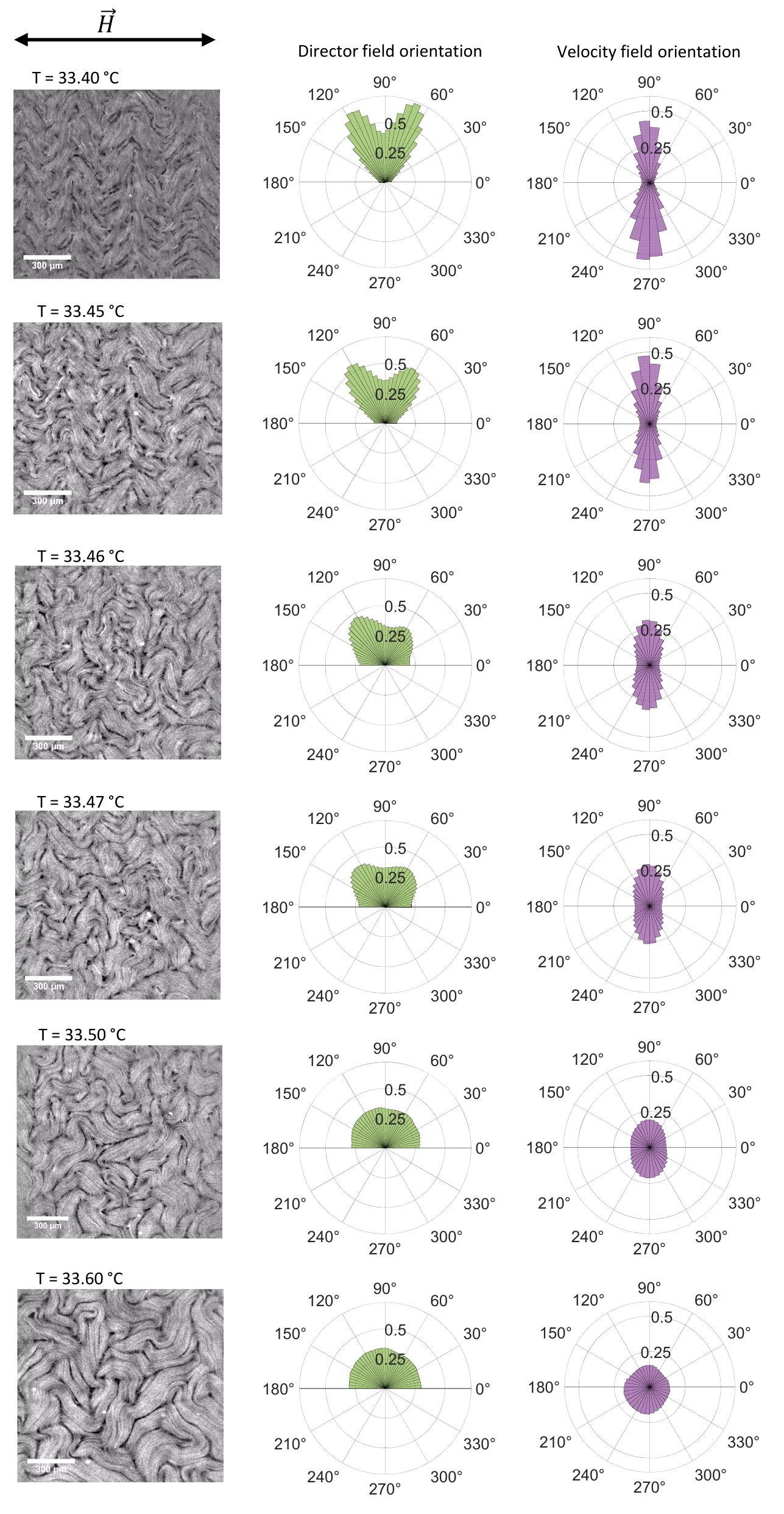}
	\caption{Alignment transition of the active nematic. The magnetic field is along the horizontal direction. Left column corresponds to fluorescence micrographs. Center column is the average distribution of director field orientations (constrained to the range $0^{\circ}$ until $180^{\circ}$). Right column is the average distribution of flow velocity orientations. Inner labels of polar plots indicate relative presence of each orientation. Scale bars, $300 \mu$m.}
	\label{sfig:alignment}
\end{figure}

\begin{figure}[t]
	\centering
	\includegraphics[width=0.5\linewidth]{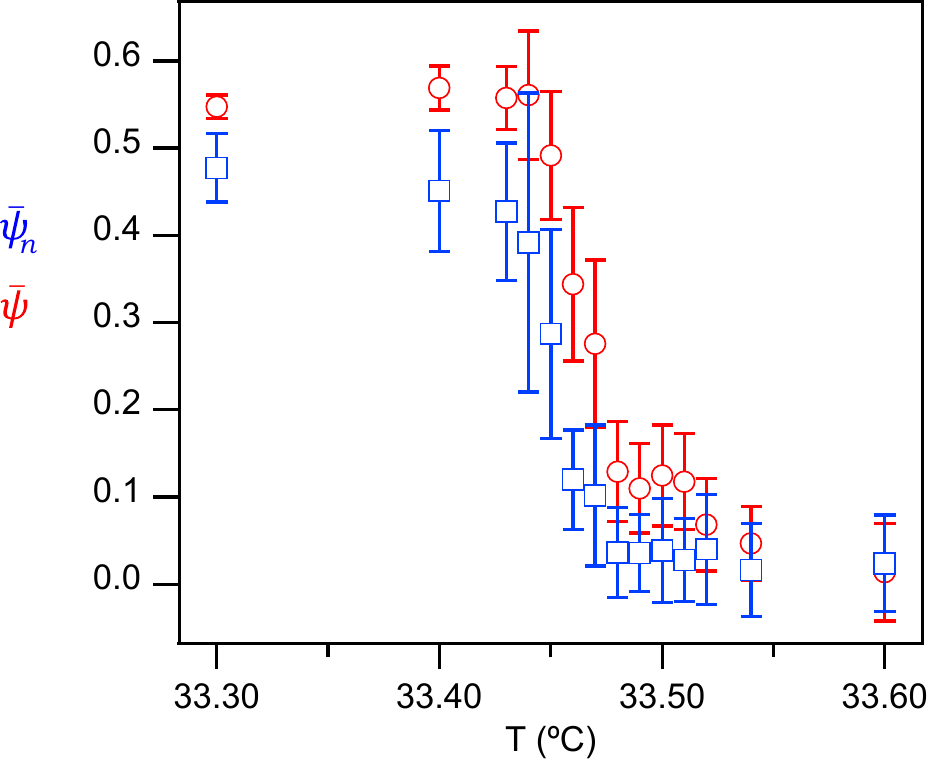}
	\caption{Evolution of the velocity ($\circ$) and director ($\square$) orientation order parameter of the active nematic averaged in space and time, as defined in the text. Temperature is changed stepwise from $33.60^{\circ}$C down to $33.30^{\circ}$C. Error bars correspond to the standard deviation of the mean order parameter at each temperature. }
	\label{Sfig:OP_transition}
\end{figure}

\begin{figure}[h]
	\centering
	\includegraphics[width=0.5\linewidth]{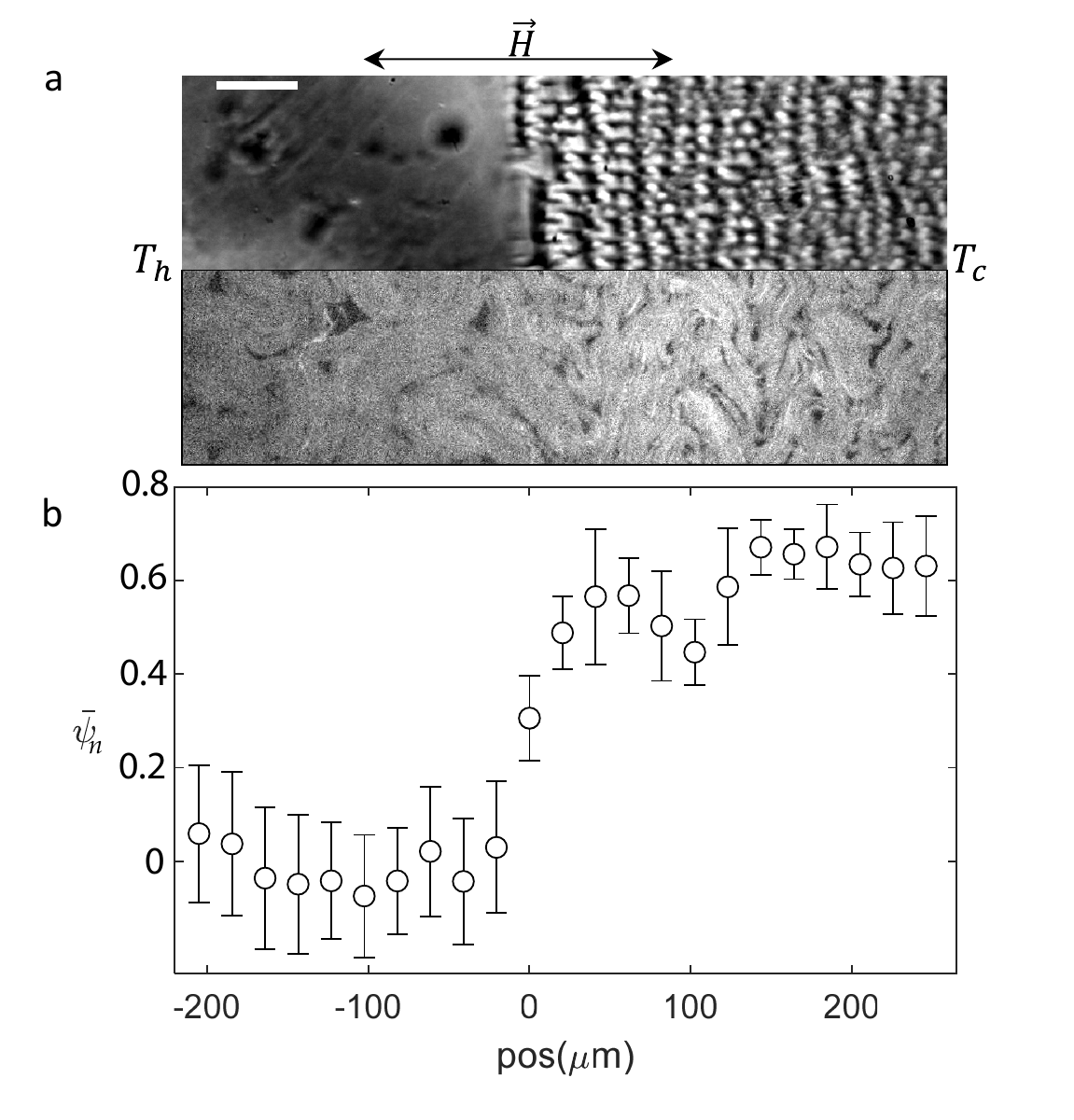}
	\caption{(a) Combined polarizing micrograph (top half) showing the state of 8CB and fluorescence micrograph (bottom half) showing the state of the AN in a horizontal temperature gradient (cold to the right, hot to the left), parallel to the magnetic field. (b) Orientational field order parameter as a function of position inside the temperature gradient. The N-SmA boundary in 8CB is set at $x=0$. Scale bar, $100 \mu$m.}
	\label{sfig:gradient}
\end{figure}

\begin{figure}[h]
	\centering
	\includegraphics[width=0.5\linewidth]{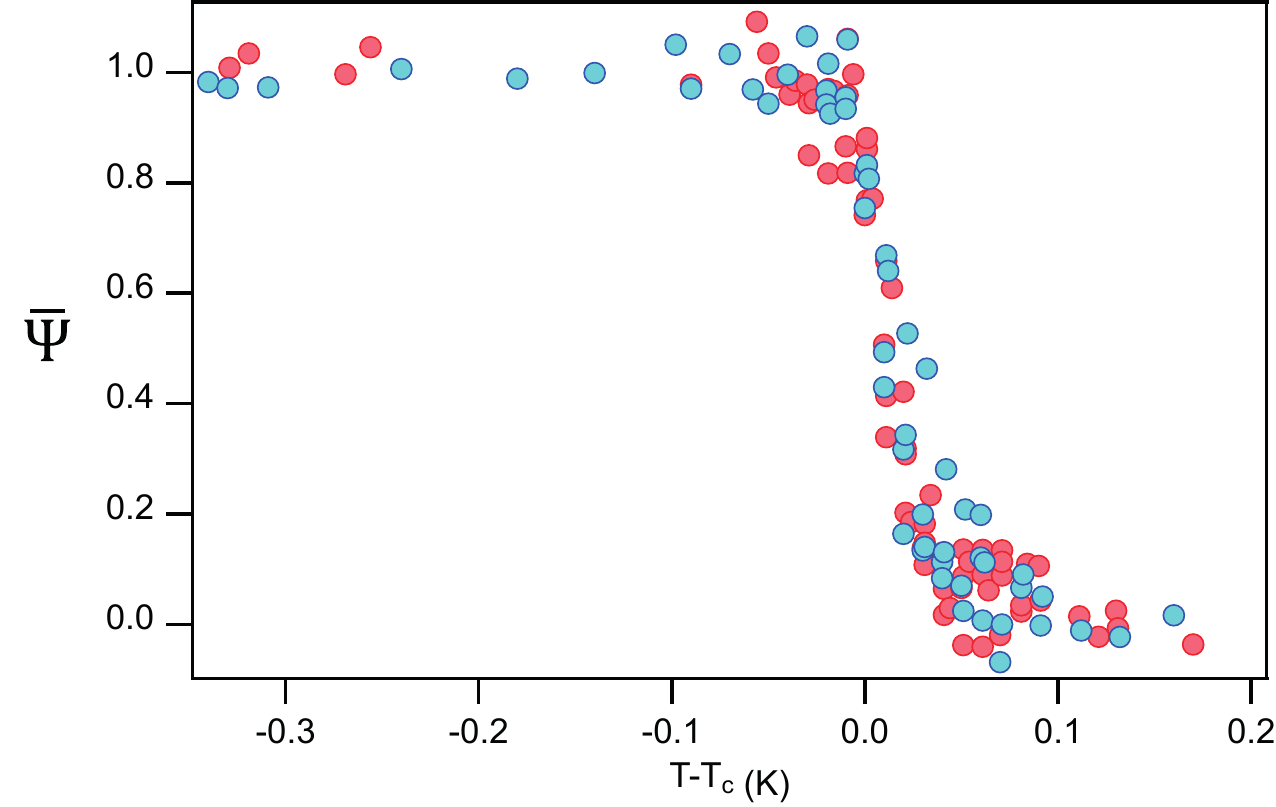}
	\caption{Evolution of the normalized director orientation order parameter $\Psi = (\bar{\psi}_n-\bar{\psi}_{n,h})/(\bar{\psi}_{n,c}-\bar{\psi}_{n,h})$, where $\bar{\psi}_n$ has been defined in the text, and $\bar{\psi}_{n,c}$ and $\bar{\psi}_{n,h}$ are, respectively, the limiting values of $\bar{\psi}_n$ for low (mesogen in SmA phase) and high (mesogen in nematic phase) temperature. Red symbols correspond to experiments performed under heating temperature ramps while blue symbols correspond to cooling ramp. The temperature axis in all data sets has been shifted with respect to $T_c$, the temperature at which the change in $\bar{\psi}_n$ is the steepest.}
	\label{sfig:OP4}
\end{figure}

\begin{figure}[h]
	\centering
	\includegraphics[width=0.9\linewidth]{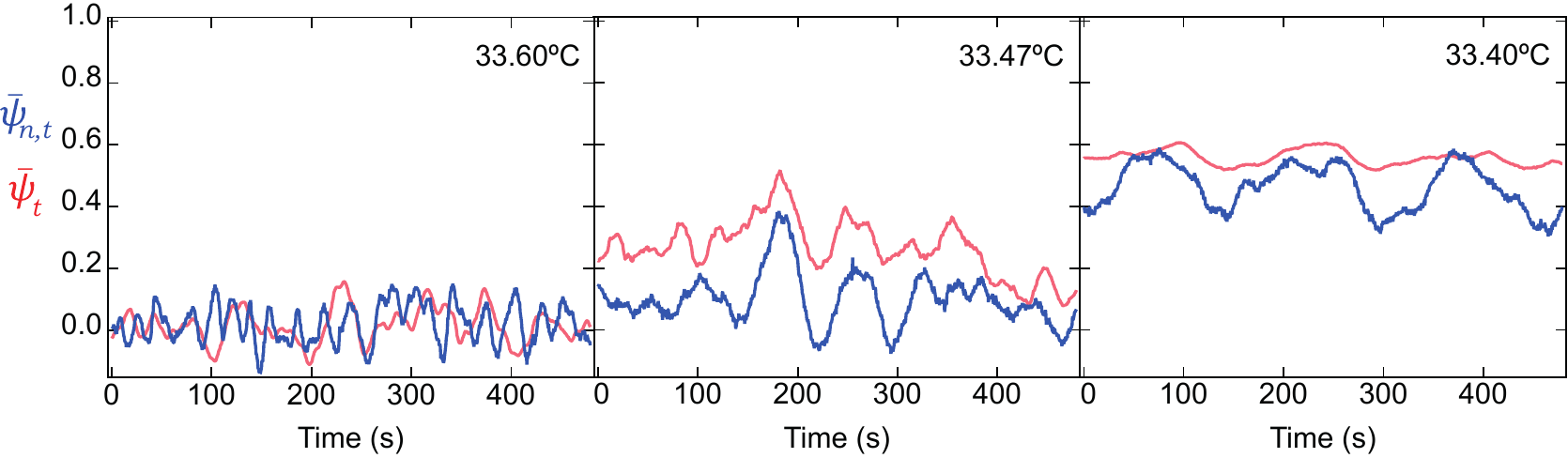}
	\caption{Comparing dynamics of the spatially-averaged velocity ($\bar{\psi}_t$) and director ($\bar{\psi}_{n,t}$) orientation order parameter of the AN for different temperatures.}
	\label{sfig:oscillations}
\end{figure}

\begin{figure}[h]
	\centering
	\includegraphics[width=\linewidth]{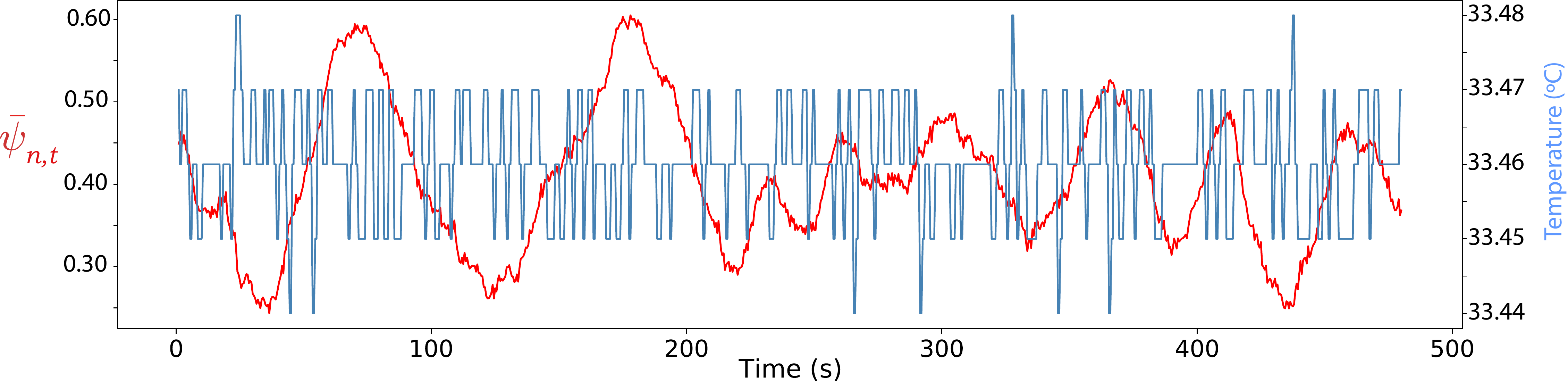}
	\caption{Comparing the oscillations in the spatially-averaged order parameter of the AN director field with the temperature fluctuations of the controller. Experiment performed at $33.46\,^{\circ}$C.}
	\label{sfig:Tfluctuations}
\end{figure}

\begin{figure}[h]
	\centering
	\includegraphics[width=0.8\linewidth]{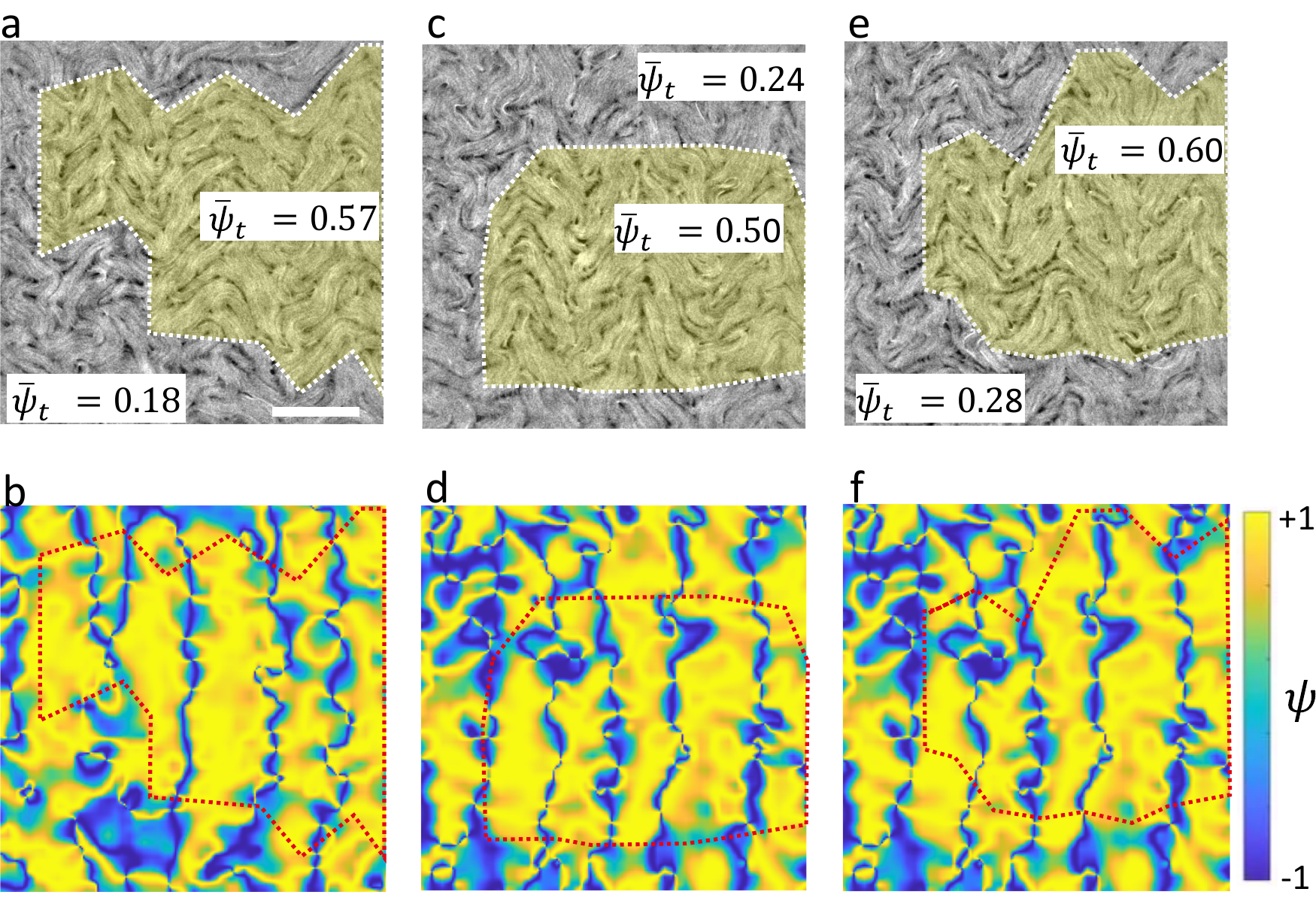}
	\caption{Spatial coexistence between aligned and disordered regions of the AN at $33.45\,^{\circ}$C obtained at three different times. a,c,e: fluorescence micrographs. b,d,f: map of the velocity order parameter $\psi=(2\sin^2(\theta_{\hat{v},\hat{H}})-1)$. In a,c,e, the average instantaneous velocity order parameter in the ordered (shaded) and in the disordered region are shown. Scale bar is 300 $\mu$m.}
	\label{sfig:coexistence}
\end{figure}

\clearpage
\onecolumngrid

\setcounter{figure}{0}

\renewcommand{\figurename}{Movie }

\begin{center}
{\bf Supplementary Movie Captions}
\end{center}

\vspace{1cm}

\begin{figure}[h]
	\centering
	\includegraphics[width=0.6\linewidth]{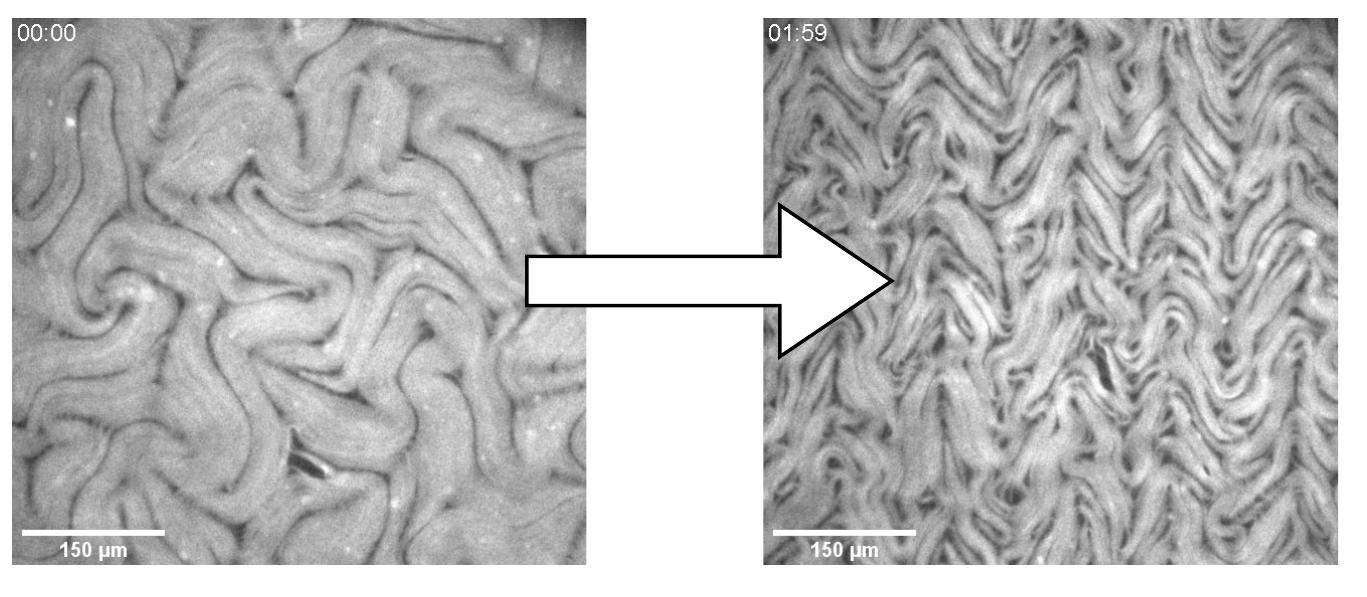}
	\caption{Transition from the turbulent to the aligned state in an active nematic during a continuous temperature ramp from $T = 33.50 ^{\circ}$C down to $T = 33.30 ^{\circ}$C.}
	\label{smov:transition}
\end{figure}

\begin{figure}[h]
	\centering
	\includegraphics[width=0.9\linewidth]{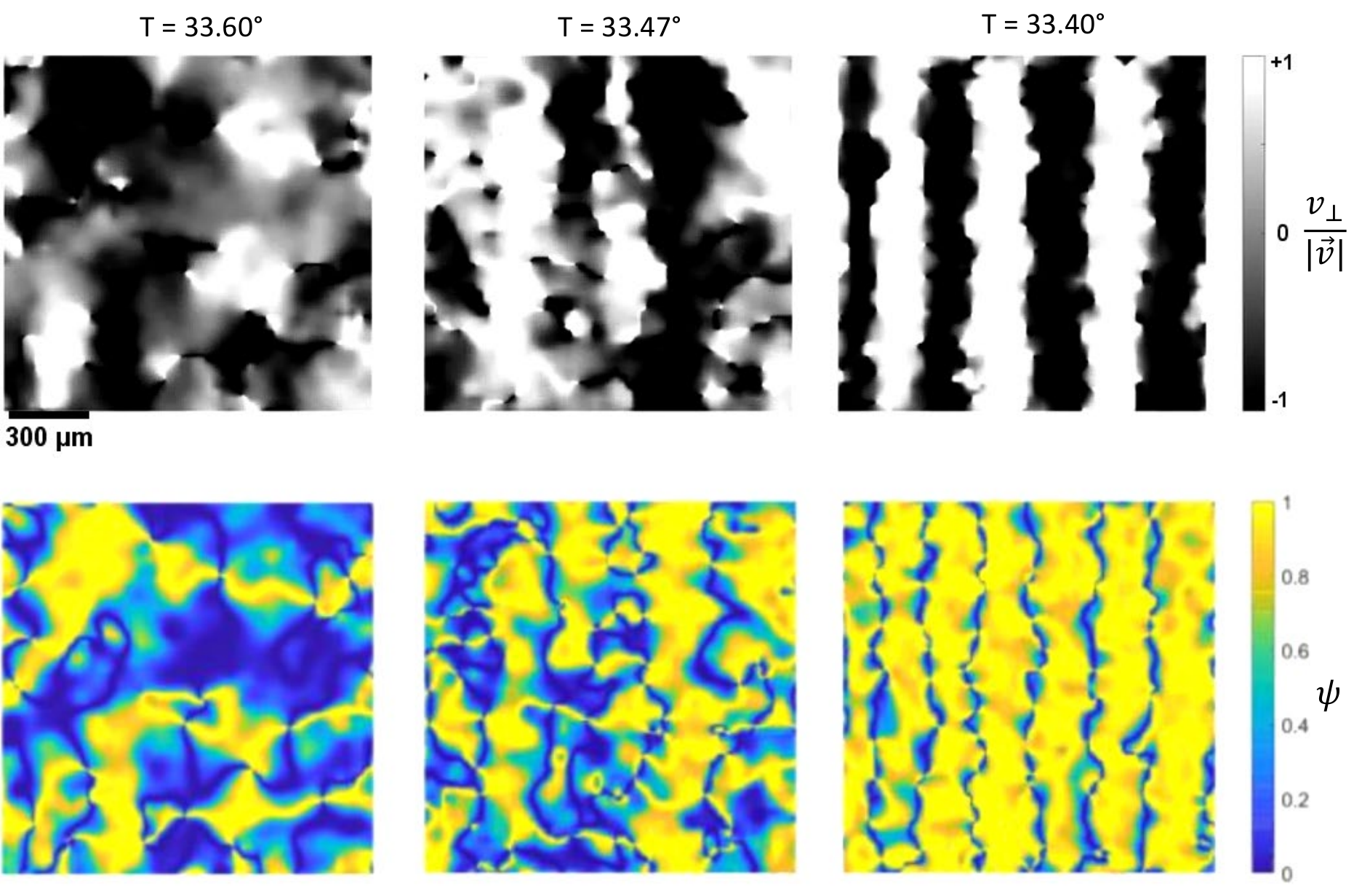}
	\caption{Velocity orientation in the active nematic at different constant temperatures, in the presence of a 1 T magnetic field. The material is interfaced with 8CB, which has the nematic to SmA transition at around $33.5 ^{\circ}$C. Top row: normalized velocity component perpendicular to the magnetic field. Bottom row: velocity orientation order parameter. }
	\label{smov:velocities}
\end{figure}

\begin{figure}[h]
	\centering
	\includegraphics[width=0.7\linewidth]{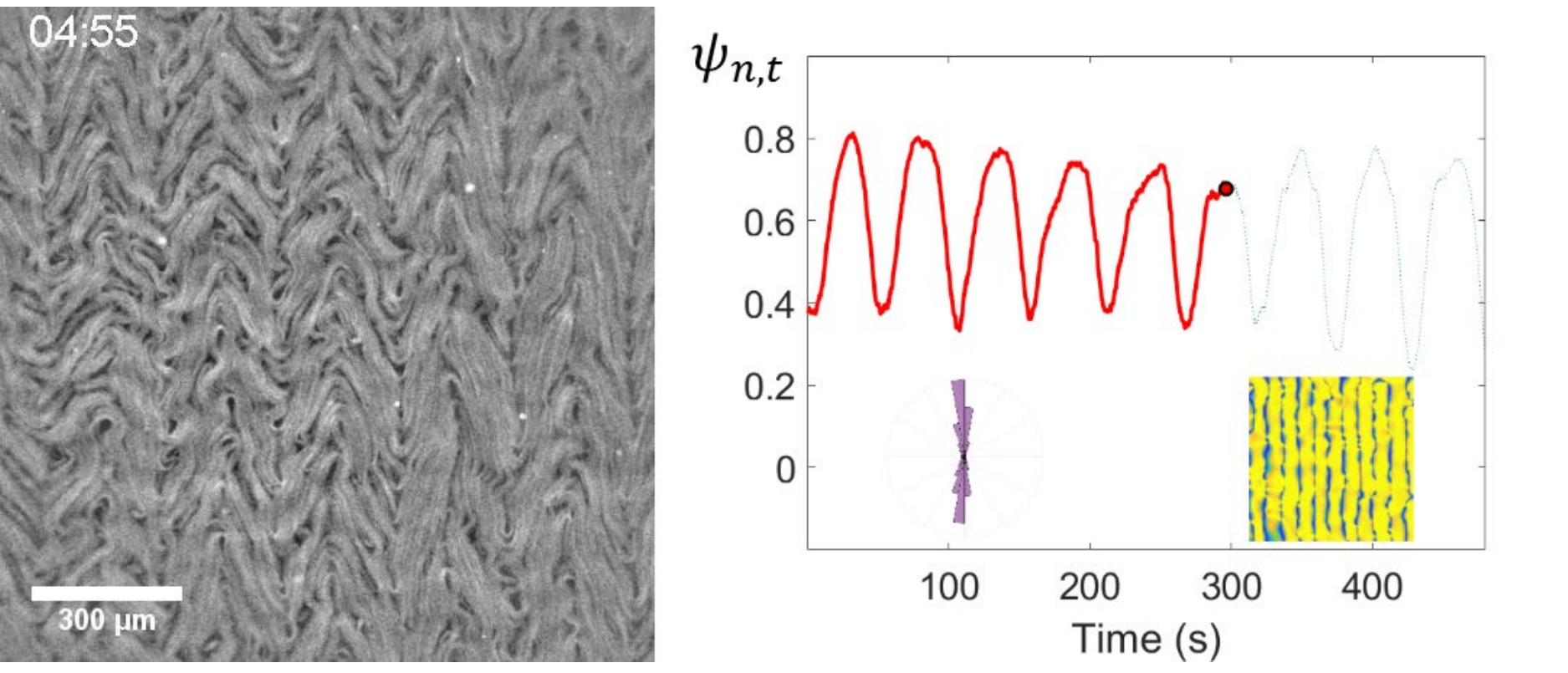}
	\caption{Oscillations of the order parameter due to transversal instabilities in the aligned state. ATP concentration is 1.37 mM, 10 times larger than the one used in this work, where these oscillations are dampened. Left: fluorescence micrographs. Right: spatially-averaged director order parameter, $\bar{\psi}_{n,t}$. Insets: Instantaneous distribution of flow velocity orientations and map of the velocity orientation order parameter.}
	\label{smov:transinst}
\end{figure}

\begin{figure}[h]
	\centering
	\includegraphics[width=0.7\linewidth]{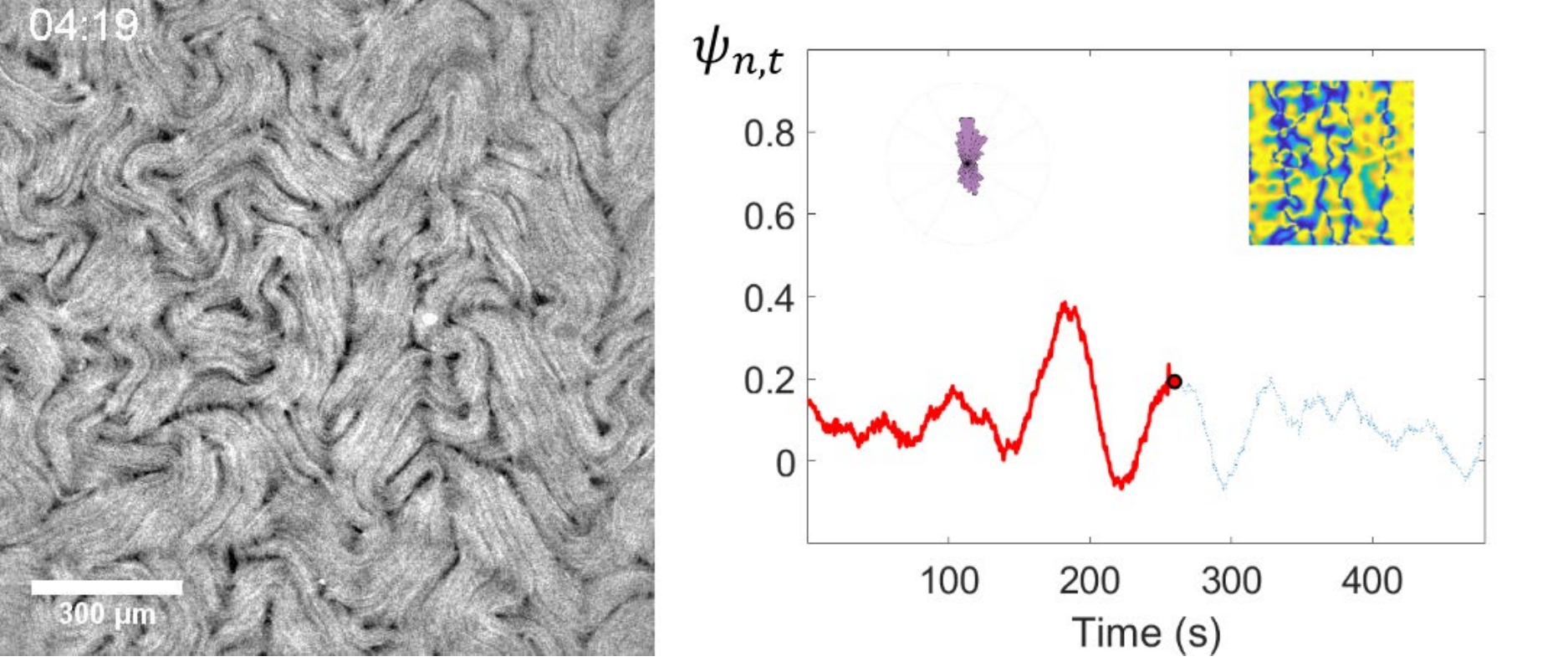
}
	\caption{Intermittency in the order parameter during the transition into the aligned state. Left: fluorescence micrographs. Right: spatially-averaged director order parameter, $\bar{\psi}_{n,t}$. Insets: Instantaneous distribution of flow velocity orientations and map of the velocity orientation order parameter.}
	\label{smov:intermittency}
\end{figure}


\end{document}